\documentclass[a4paper,11pt,nohyper]{JHEP3}
\usepackage{amsmath,latexsym,amssymb}
\usepackage{graphicx,euscript}
\usepackage{array,calc}
\usepackage{psfrag}
\usepackage[latin1]{inputenc}
\usepackage{subfigure}
\usepackage{nicefrac}
\usepackage{bbm}
\usepackage[stable]{footmisc}
\usepackage{fancybox}
\usepackage{cite}

\usepackage[]{graphicx}

\usepackage{comment}

\newcommand{\arXividhepth}[1]{\href{http://arxiv.org/abs/#1}arXiv:{\tt #1} [hep-th]}

\def\d{\text{d}}

\def\slashchar#1{\setbox0=\hbox{$#1$}           
\dimen0=\wd0                                 
\setbox1=\hbox{/} \dimen1=\wd1               
\ifdim\dimen0>\dimen1                        
\rlap{\hbox to \dimen0{\hfil/\hfil}}      
#1                                        
\else                                        
\rlap{\hbox to \dimen1{\hfil$#1$\hfil}}   
/                                         
\fi}

\def\Re           {{\rm Re\hskip0.1em}}
\def\Im           {{\rm Im\hskip0.1em}}

\def\be{\begin{equation}}
\def\ee{\end{equation}}
\def\bea{\begin{eqnarray}}
\def\eea{\end{eqnarray}}
\newcommand\bbone{\ensuremath{\mathbbm{1}}}
\newcommand{\ul}{\underline}
\def\bseq{\begin{subequations}}
\def\eseq{\end{subequations}}

\def\f4{\zeta}
\newcommand{\psone}{\Phi_1}
\newcommand{\pstwo}{\Phi_2}
\newcommand{\NSB}{{B}}
\newcommand{\warp}{{Z}}
\newcommand{\bindex}{{B}}

\newcommand{\cof}{e}
\newcommand{\fourdspinor}{\psi}
\newcommand{\comp}{Y}

\def\del {\partial}

\def\ka {K\"ahler}
\def\del {\partial}

\def\d {{\rm d}}

\def\calc         {{\cal C}}
\def\cald         {{\cal D}}

\def\calg         {{\cal G}}

\def\calh         {{\cal H}}
\def\cali         {{\cal I}}

\def\calk         {{\cal K}}

\def\calm         {{\cal M}}
\def\caln         {{\cal N}}

\def\calt         {{\cal T}}

\def\calz         {{\cal Z}}

\title{Domain walls from ten dimensions}

\author{Michael Haack$^1$, Dieter L\"ust$^{1,2}$, Luca Martucci$^1$ and Alessandro Tomasiello$^{3,4}$ \\

\begin{itemize}
\item[${}^1$]  Arnold-Sommerfeld-Center for Theoretical Physics\\
Ludwig-Maximilians-Universit\"at M\"unchen\\
Theresienstra\ss e 37, 80333 M\"unchen, Germany
\item[${}^2$]  Max-Planck-Institut f\"ur Physik\\
F\"ohringer Ring 6, 80805 M\"unchen, Germany
\item[${}^3$]  Jefferson Physical Laboratory, Harvard University,
Cambridge, MA 02138, USA
\item[${}^4$]  Universit\`a di Milano--Bicocca
and INFN, sezione di Milano--Bicocca,
I-20126 Milano, Italy
  \end{itemize}
}

\abstract{We write down the general conditions for ${\cal N}=1$ supersymmetric four--dimensional domain walls, deriving them from a ten--dimensional point of view using generalized complex geometry. In cases where the compactification allows for a truncation to a finite number of fields, we make contact with a four--dimensional effective description. In the context of the AdS/CFT correspondence, the equations can be applied to renormalization--group flows of three--dimensional field theories. We allow for the presence of explicit brane sources and show how supersymmetry restricts their location in a natural way.}

\preprint{MPP-2009-54\\
LMU--ASC 19/09}


\begin{document}
\setcounter{footnote}{0}
\renewcommand{\thefootnote}{\arabic{footnote}}
\setcounter{section}{0}
\section{Introduction}
\label{introduction}

There are several reasons to study domain wall solutions. In four dimensions, a time--honored application is to compute tunneling--rates of vacua in a theory with many vacua, such as string theory. The recent revival of AdS$_4$/CFT$_3$ (starting from \cite{Aharony:2008ug}) offers another reason: renormalization--group (RG) flows in the field theory can holographically be thought of as domain walls in the bulk. This is well known since the early work of \cite{Girardello:1998pd,Distler:1998gb,FGPW}. 

These solutions are typically found using an effective four--dimensional supergravity action, see for instance 
\cite{Cvetic:1992bf,Behrndt:2001mx,Louis:2006wq,Ceresole:2006iq}.
This approach, however, has its obvious limitations. Sometimes, a new AdS$_4$ vacuum is found directly in ten dimensions, for which an effective four--dimensional description is not available yet. It could also run into technical trouble when the effective theory in four dimensions is not a consistent truncation of the theory in ten dimensions, since it is then not guaranteed that the four--dimensional solution found in this way really represents a solution in ten dimensions. 

In order to alleviate these problems, in this paper we study the equations governing supersymmetric domain walls directly in ten dimensions.\footnote{Recently, AdS$_4$ domain wall solutions in ten dimensions, including (partly) smeared brane sources, were constructed in \cite{klpt,klt}.} We make no assumptions on the internal space, and some reasonable ones on the allowed fluxes, on top of imposing the symmetries of the problem. In this way, we extend the analysis of \cite{Mayer:2004sd,Louis:2006wq} (see also \cite{smyth} for related work). We find a system of equations, (\ref{flow1c})--(\ref{algconst2}), that involve both radial derivatives and exterior differentials in the internal space. The latter can be a Calabi--Yau manifold or a more general space. Our focus is on the latter case: even though there are many supersymmetric AdS$_4$ solutions in string theory, CFT$_3$ interpretations have been found so far only for non--Calabi--Yau solutions. 

Assuming a certain truncation ansatz, presented in section \ref{truncation}, we can actually show that the ten--dimensional flow equations (\ref{flow1c})--(\ref{algconst2}) can be rewritten as the four--dimensional flow equations in an appropriate supergravity theory. As one expects from the results of \cite{Cvetic:1992bf,Ceresole:2006iq}, they become the gradient flow for a certain ``generating function'' ${\cal C}$ (see (\ref{flowc})). This function has different interpretations. Evaluated at  the AdS$_4$ solutions connected by the domain wall, it just gives their corresponding cosmological constants, but it can also be seen as a holographic c-function. We will elaborate on this point in section \ref{ctheorem}.

The truncation ansatz includes certain internal spaces (among them the cosets $S^6$ and $\mathbb{CP}^3$), for which AdS$_4\times{\cal M}_6$ vacua have been found in massive IIA supergravity, without any smeared sources \cite{behr,tomtwistor,klt}. In these cases, one can find effective field theories \cite{kp,ck,paltihouse,Caviezel:2008ik} which have been shown (at the bosonic level \cite{ck}) to be consistent truncations of massive IIA supergravity. 
These examples, presented in section \ref{explicit}, provide a useful playground for some preliminary studies of the equations, and they might lead to their most immediate applications in the future. Two (related) possibilities that spring to mind are flows connecting different AdS$_4$ vacua\footnote{In the maximal ${\cal N}=8$ theory, such flows were discussed in \cite{Ahn:2000aq,Bobev:2009ms}.}, or brane solutions whose near--horizon limits reproduce given AdS$_4$ vacua. Such brane solutions are easy to find for AdS$_5$ vacua, but are not known for any AdS$_4$ vacuum with Romans mass, for example.

Until now, we did not find any explicit (numerical) solutions to the flow equations. 
One challenge is the following. The AdS$_4$ vacua are expected to be critical points of the
function ${\cal C}$ that generates the flow.  
However, in the simplest cases we have studied, ${\cal C}$ has 
an indefinite Hessian. For $S^6$, for example, we kept four fields in our explicit analysis. The corresponding Hessian has two positive and two negative eigenvalues. For this reason, the AdS vacua are fixed points of the flow but they are not quite attractors.
They are ``semi--attractors'', in the sense that the flow is attractive in two directions and repulsive in the other two. In other words, if one starts at some finite value of the radial coordinate, one can reach the AdS vacua in the infrared, but only after fine--tuning two boundary conditions out of four. This is reminiscent of other flows discussed in the literature, see for instance \cite{FGPW, Bobev:2009ms}.

We also include explicit D--brane sources in our analysis. This is usually not done in RG flow solutions. We include them for several reasons. First of all, many of the supersymmetric AdS$_4$ solutions known so far have non--zero Romans mass $F_0$. Such a flux is sourced by a D8--brane. Unlike for other branes, the harmonic function in its gravity solution does not diverge on it. It just creates a discontinuity in the first derivatives of some fields which is easy to incorporate in the flow equations. 

Second, having the possibility to change the flux parameters during the flow by including explicit D--brane sources widens the scope for finding solutions interpolating between different supersymmetric AdS$_4$ minima. For instance, in the case of nearly K\"ahler reductions, there is only a single supersymmetric AdS$_4$ vacuum for a {\it fixed} set of fluxes \cite{ck}.

Finally, explicit inclusion of branes could allow us in the future to find brane solutions with assigned AdS near--horizon limits, as mentioned earlier. For example, for reasons similar to the ones discussed in \cite{Chen:1999bf} (see also \cite{Lee:2008ha,Camps:2008hb}) one expects the existence of a solution with D2-- and D8--branes (with non--zero $B$--field) separated by a finite distance. This would be reminiscent of the existence of multi--center black hole solutions in supergravity \cite{denef}.\footnote{We thank F.~Denef for discussions on this point.} Unless the D8--branes were spherically shaped around the D2--branes, the functions in such a solution would depend on two radial variables (the distances from the D2-- and the D8--branes). This makes finding an explicit solution a difficult problem. Nevertheless, we find qualitatively that the positions of the D--brane sources are fixed by supersymmetry, in a way reminiscent of how several black holes can be in equilibrium with each other at finite distance. This can be seen to follow both from the bulk supersymmetry conditions, and from the conditions that the D--brane should be calibrated.

The organization of the paper is as follows. In section \ref{calibrations} we derive the general supersymmetry conditions for ten-dimensional flows having ${\cal N}=1$ supersymmetry and three-dimensional Lorentz-invariance. Section \ref{4Dint} paves the way for the four--dimensional effective description of the flow, by introducing the relevant K\"ahler-- and superpotential. In section \ref{calibrations}, however, we do not yet perform a truncation to a finite number of four-dimensional fields. That truncation is done in the following sections.  In section \ref{truncation} we apply our general flow equations to the case of SU(3) structure manifolds and in section \ref{explicit} we specialize further to the cases of coset spaces and nearly K\"ahler manifolds. In those cases, we can make contact with the general form of flow equations derived within the context of four--dimensional ${\cal N}=1$ supergravity \cite{Cvetic:1992bf,Ceresole:2006iq}. Our formulas for the K\"ahler-- and superpotential are not new in these cases: they can be found in the effective theories we mentioned earlier \cite{kp,ck,paltihouse,Caviezel:2008ik}, and they could also be obtained by specializing the general results of \cite{granasup,grimmsup,effective,casbil}. However, deriving the flow equations from the much more general ten-dimensional equations of section \ref{calibrations} simplifies a lift to ten dimensions and also allows to discuss the inclusion of explicit D--brane sources into the flow. This latter point is further discussed in section \ref{inclusion}, where we also comment on the effect of branes on the holographic c-theorem discussed in the literature. Finally, we collected some more technical parts of the calculation and a short introduction into some basics of generalized geometry in the appendix.


\section{${\rm G}_2\times{\rm G}_2$ structure and flow equations: the general case}
\label{calibrations}

In this section we analyze the conditions for supersymmetric domain walls in four dimensions. 
Specifically, we are looking for general type II supersymmetric backgrounds preserving $\caln=1$ super-Poincar\'e symmetry in three dimensions, i.e.~two real supercharges. 

Let us consider, then, the space $\mathbb{R}^{1,2}\times {\cal M}_7$, with metric 
\bea
\d s^2=e^{2\hat{A}}\d x^\mu\d x_\mu+\d s^2_7\ .
\eea
As a consequence of the Poincar\'e symmetry, we can decompose the total RR polyform as
\bea
F_{\rm tot}=F+ \d{\rm vol}_3 \wedge *_7 \lambda(F)\ ,
\eea
where $\d{\rm vol}_3 = e^{3\hat{A}} \d x^1 \wedge \d x^2 \wedge \d x^3$, and $\lambda$ is multiplication by a sign, as defined in 
(\ref{eq:lambda}). The only flux that we do not consider is $H$ along ${\Bbb R}^{1,2}$.

The ten--dimensional Killing spinors associated to the unbroken ${\cal N}=1$ supersymmetry can be expressed in terms of two seven--dimensional Majorana spinors $\chi_1$ and $\chi_2$ -- see appendix \ref{notconv}. These can be used to construct two real polyforms $\Psi_{1,2}$, defined as follows. Take 
\begin{equation}\label{eq:g2g2}
	 \Psi = \frac 8{|\chi|^2}\,\chi_1 \otimes \chi_2^{\dagger}\ .
\end{equation}
As usual, here we have used the Clifford map to identify bispinors with differential forms; $|\chi|^2$ is the norm of both $\chi_1$ and $\chi_2$ (see (\ref{eq:chi2})).  
Take, then, its even and odd part, $\Psi = \Psi_+ + i \Psi_-$.  $\Psi_\pm$ turn out to be real. Now, in IIA we take $\Psi_1=\Psi_+$, and $\Psi_2= \Psi_-$, in IIB we take $\Psi_1=\Psi_-$, and $\Psi_2= \Psi_+$. The polyforms $\Psi_1$ and $\Psi_2$ define a ${\rm G}_2\times{\rm G}_2$ structure \cite{witt}. Notice that $\Psi_1$ and $\Psi_2$ satisfy the normalization condition
\bea
\langle \Psi_1, \Psi_2\rangle=8 \,\d \text{vol}_7\ ,
\eea
with $\langle\,,\rangle$ is the seven-dimensional Mukai pairing (see (\ref{mukai})).

One can then show that the conditions for unbroken ${\cal N}=1$ supersymmetry are equivalent to
\bea \label{sc2}
&\d_H(e^{3\hat{A}-\phi}\Psi_1)=-e^{3\hat{A}} *_7\lambda(F)\ ,\qquad  \d_H(e^{2\hat{A}-\phi}\Psi_2)=0&\ ,\\ 
&\langle \Psi_2, F\rangle=0\ .& 
   \label{algconst1}
\eea
The calculations are similar to the ones in \cite[App.~A]{scan}.\footnote{Notice that $e^{3\hat{A}-\phi}\Psi_1$ and $e^{2\hat{A}-\phi}\Psi_2$  are generalized calibrations (in the sense of \cite{gencal,lucajarah}) for D--branes which are space-filling and string-like in $\mathbb{R}^{1,2}$, respectively.} 
These equations were also considered in \cite{jeschek-witt-shame} in the case without warping and in \cite[App.~B]{adsbranes} for the AdS$_4$ case.

Let us now assume that the internal seven--dimensional space ${\cal M}_7$ can actually be seen as a foliation over ${\Bbb R}$, whose generic leaf is a six--dimensional space $\calm_6$. We parameterize (an open subset of) $\mathbb{R}$ by the coordinate $r$. In order to facilitate a four--dimensional interpretation of the flow equations later on, it is convenient to use the following parameterization of the ten--dimensional metric
\bea\label{10Dmansatz}
\d s^2=e^{2\warp}(e^{2A}\d x^\mu\d x_\mu+\d r^2)+\d s^2_6\ ,
\eea
where $A$ depends just on $r$, while $\warp$ can depend on the internal coordinates $y^m$ on $\calm_6$ as well. The ten--dimensional Majorana--Weyl Killing spinors $\epsilon_{1,2}$ can now be split as follows
\bea
\epsilon_{1,2}=\psi\otimes\eta_{1,2}+\ \text{c.c.}\ ,
\eea
where $\eta_{1,2}$ are chiral spinors in six dimensions,  while $\fourdspinor$ is a constant four--dimensional chiral spinor satisfying the projection condition
\bea\label{firstproj}
\gamma_{\ul r}\fourdspinor=\fourdspinor^*\ ,
\eea 
where $\gamma_{\ul r}$ is the gamma matrix along $r$, in frame indices. One can then introduce on $\calm_6$ two complex polyforms $\Phi_{1,2}$ defined by
\bea \label{6dpurespin}
\psone=e^{3Z-\phi} \frac{8}{i|\eta|^2}\eta_1\otimes\eta^\dagger_2\ , \qquad \pstwo=e^{3Z-\phi} \frac{\pm 8}{i|\eta|^2}\eta_1\otimes\eta^T_2\ ,
\eea 
where $|\eta|^2\equiv \eta^\dagger_1\eta_1=\eta_2^\dagger\eta_2$. One can show, then, that $\Phi_{1,2}$ are ${\rm O}(6,6)$ pure spinors defining an SU(3)$\times$SU(3) structure on ${\cal M}_6$ and satisfy the normalization condition
\bea
i\langle\Phi_1,\bar\Phi_1\rangle=i\langle\Phi_2,\bar\Phi_2\rangle=8e^{6\warp-2\phi}\d\rm{vol}_6\ ,
\eea
which has been chosen so that the resulting supersymmetry conditions (see below) are not cluttered with factors of $e^{3Z-\phi}$. $\Phi_{1,2}$ and $\Psi_{1,2}$ are related by
\bea \label{eq:PsiPhi}
\Psi_1= e^{-3Z+\phi}(e^\warp \d r\wedge \Re \psone+\Re\pstwo) \ , \qquad \Psi_2=e^{-3Z+\phi}(\Im \psone+e^\warp \d r\wedge \Im\pstwo) \ .
\eea

Let us now split the fluxes on ${\cal M}_7$ as follows
\begin{equation}
	\begin{split}
		F\,\rightarrow\,& F+\d r\wedge F_r\ , \\
		H\,\rightarrow\,& H +\d r\wedge H_r \ , \label{fluxansatz}
	\end{split}
\end{equation}
where, on the right-hand side, $F, F_r, H$ and $H_r$ have legs only along $\calm_6$.  Similarly, we split the D--brane currents according to
\bea
j\rightarrow j+\d r\wedge j_r\ .
\eea

In the following, we assume that the NS three--form flux $H$ is exact, $H=\d B$, and that 
$B$ has only internal indices; this assumption is not essential and can be easily relaxed. We mostly work in the twisted picture, obtained by substituting (polyform) $\rightarrow$ (polyform)$^\bindex\equiv e^\NSB$(polyform). Defining $*_\bindex\equiv e^\NSB*_6\lambda e^{-\NSB}$, \eqref{sc2} can be written as the following system of flow equations\footnote{To obtain the corresponding equations in the untwisted picture, one has to replace the twisted by the untwisted polyforms and further make the replacements: $*_\bindex \rightarrow *_6 \lambda, \d \rightarrow \d_H$ and $\partial_r \rightarrow \partial^H_r\equiv \partial_r+H_r \wedge$. In this form the equations would be valid even if $H$ were not exact. \label{Hr}}
\bea 
\d(e^{\warp}\Re\psone^\bindex)&=&e^{4\warp}*_\bindex F^\bindex+\partial_r \Re\pstwo^\bindex
+3 \Re \pstwo^\bindex \partial_r A\ ,\label{flow1c}\\
\d(\Re\pstwo^\bindex)&=&-e^{2\warp}*_\bindex F^\bindex_r\ ,\label{flow2c}\\
\d(e^{-\warp}\Im\psone^\bindex)&=&0\ ,\label{flow3c}\\
\d(\Im\pstwo^\bindex)&=&\partial_r(e^{-\warp}\Im\psone^\bindex)+2 e^{-\warp}\Im \psone^\bindex \partial_r A \label{flow4c}\ ,
\eea
while the constraint (\ref{algconst1}) becomes
\bea\label{algconst2}
\langle \Im\pstwo^\bindex,F^\bindex \rangle+e^{-\warp}\langle \Im\psone^\bindex,F_r^\bindex \rangle=0\ .
\eea
Similar equations were derived independently by the authors of \cite{smyth}.
If one considers the case in which $A$ is linear in $r$, $\Phi_{1,2}^\bindex$, $Z$ and
$F^\bindex$ are indepedent of $r$ and $F_r^\bindex=0$, one recovers the conditions for a
supersymmetric AdS$_4$ vacuum \cite{scan} (see also \cite{lt} in the SU(3)
structure case in IIA).
The RR equations of motion simply follow from \eqref{flow1c} and \eqref{flow2c}
and the Bianchi identities are 
\bea\label{splitBIb}
\d F^\bindex=-j^\bindex\ , \qquad \d F^\bindex_r-\partial_r F^\bindex= j^\bindex_r\ .
\eea
As for the NS flux $H$, its Bianchi identity $\d H=0$ is taken care of by our simplifying assumption that $H$ is exact. Its equation of motion can be seen to follow from the supersymmetry equations and the Bianchi identities \cite{kt,adsbranes,Lust:2008zd} -- as do the equations of motion for the metric and dilaton \cite{Gauntlett:2002fz,lt,Gauntlett:2005ww}.

The D--brane currents are defined as follows. For a D--brane wrapped around a cycle $\Sigma$ of ${\cal M}_7$ (which might or might not include the $r$-direction) and with world-volume flux $F^{({\rm wv})}$, the current $j^\bindex$ on ${\cal M}_7$, that we introduced above, is defined by demanding that for any polyform $\omega$ on ${\cal M}_7$ one has
\bea
\int_{{\cal M}_7} \langle \omega, j^\bindex \rangle = \int_\Sigma \omega \wedge e^{F^{({\rm wv})}}\ , 
\eea
i.e.\ $j^\bindex = e^{-F^{({\rm wv})}} \wedge \delta(\Sigma)$, where $\delta(\Sigma)$ has legs transversal to $\Sigma$. We note as an aside that the contribution of O--planes to the currents requires including some normalization factors, which will not, however, play any role in the following.

We can split
\bea \label{splitF}
F^\bindex=F^{(0)}+\d C\ ,
\eea 
where $F^{(0)}$ is some background flux, which is constant away from sources (whose effect will be considered in
section \ref{inclusion}) and $C$ is the dynamical RR potential (with only internal indices). We further assume that there is no background for $F^\bindex_r$, which implies
\bea \label{eq:frc}
F^\bindex_r=\partial_r C\ .
\eea


\section{Towards the four--dimensional effective description}
\label{4Dint}

Following \cite{effective,Martucci:2009sf}, let us also introduce the polyforms
\bea \label{calz}
\calz=\pstwo^\bindex\ , \qquad T=e^{-3 \warp}\psone^\bindex \ , \qquad \calt=\Re T-iC\ .
\eea
As we will see, $\calz$ and $\calt$ will play the role of chiral fields in the four--dimensional description, which is why we prefer to give these polyforms their own names. Notice that
\bea\label{warpf}
e^{6\warp}=\frac{\langle\calz,\bar\calz\rangle}{\langle T,\bar T\rangle}
\eea 
and, thus, the warp factor $e^\warp$ can be considered as a function of $\calz$ and $T$. Furthermore, one should keep in mind that $\Re T$ (and thus $\calt$) contains the full information about $T$  \cite{hitchin}. Hence, for fixed $F^{(0)}$, $\calz$ and $\calt$ contain the full information about the background.

In the following sections we will make use of some properties of the conformal compensator formulation of supergravity, as it allows to switch easily between different frames in four dimensions. Such a formulation possesses an invariance under local Weyl and chiral transformations.  A good review can be found in \cite{Kallosh:2000ve}. 

Let us first focus on Weyl transformations, which  transform the four dimensional metric with weight $-2$, i.e.\ $g_{4}' = e^{-2 \sigma} g_4$. As discussed in \cite{effective}, in our setting, the invariance of the four dimensional action arises because of an ambiguity in the split in \eqref{10Dmansatz}. The ten-dimensional metric is invariant under the simultaneous transformation $g_{4}' = e^{-2 \sigma} g_4$ and $Z'= Z + \sigma$. From the definition \eqref{calz} it follows that $\calz$ has Weyl--weight 3,  
\bea
\calz\rightarrow e^{3\sigma}\calz\ .
\eea
On the other hand $T$ and $\calt$ have Weyl--weight $0$.

This ambiguity can of course be fixed at will, but a natural choice is the one leading to a canonically normalized four--dimensional Einstein--Hilbert term for the unwarped four--dimensional metric $g_4$. Indeed, by a simple dimensional reduction, it is easy to see that the Einstein--Hilbert term comes with a prefactor which, expressed in terms of the polyforms (\ref{calz}), reads\footnote{We work in units in which $2\pi\sqrt{\alpha^\prime}=1$.}
\bea
\caln=\frac{i\pi}{2}\int_{\calm_6}\langle\calz,\bar\calz\rangle^{1/3}\langle T,\bar T\rangle^{2/3}\ ,
\eea
cf.\ \cite{effective} for more details. Notice that $\caln$ has Weyl--weight 2. The four--dimensional Einstein frame corresponds to the condition
\bea\label{einstcond}
\caln=M_{\rm P}^2\ .
\eea

By fixing this frame, one can see that the flow equations derived in section \ref{calibrations} imply that
\bea \label{adotgen}
\dot A=\frac{\pi}{M^2_{\rm P}}\int_{\calm_6}\langle\calz,F^{(0)}+i\d\calt\rangle =: -\calc\ .
\eea
For more details we refer to appendix \ref{flowdetails}. We defined the function $\calc$ here, which will appear again later when we discuss a brane-modified c-theorem, cf.\ sec.\ \ref{ctheorem}.
Notice that, by using \eqref{flow2c}, the condition (\ref{algconst2}) can be written as
\bea\label{algconst}
\Im\langle \calz,F^{(0)}+i\d\calt \rangle=0\ ,
\eea
which implies that indeed the r.h.s.~of (\ref{adotgen}) is real. We stress that eq.~\eqref{adotgen} holds generally, i.e.\ even with nontrivial warping $\warp$ and without truncating to a finite number of fields. 

Equation (\ref{adotgen}) for the warp factor can directly be related to a corresponding equation derived from a purely four--dimensional analysis \cite{Cvetic:1992bf,Ceresole:2006iq}. To see this, we need to pass from the superconformal formulation to the ordinary Einstein one, but first we have to explain where the chiral gauge invariance of the former comes from. In fact, the specific ten--dimensional spinorial ansatz introduced in the previous section has already broken this symmetry. One can reintroduce it by defining a new 4-6 split of the ten--dimensional spinors: 
\bea\label{newsplit}
\epsilon_{1/2} = \fourdspinor \otimes \eta_{1/2} + \text{c.c.} = \fourdspinor e^{i\vartheta/2} \otimes e^{-i\vartheta/2} \eta_{1/2} + \text{c.c.} = \fourdspinor^{(\rm new)} \otimes \eta_{1/2}^{(\rm new)} + \text{c.c.} \ .
\eea
This implies that
\bea \label{znew}
\calz^{(\rm new)}=e^{-i\vartheta}\calz\ .
\eea
Omitting the superscript ``(new)'', we see that the new $\fourdspinor$ satisfies\footnote{\label{phaseconv}Notice that in our conventions the four--dimensional chirality operator has an opposite overall sign with respect to the one used in \cite{Ceresole:2006iq}. Thus,  $e^{i\vartheta}=\mp e^{-i\theta_{\rm there}}$, where the sign ambiguity is as in \cite{Ceresole:2006iq}.}
\bea
\gamma_{\ul r}\fourdspinor= e^{i\vartheta}\fourdspinor^*\ .
\eea
In our applications, we usually assume $\vartheta=\vartheta(r)$. 

 Thus, we clearly have a chiral gauge symmetry under which $\fourdspinor\rightarrow e^{-i\alpha/2}\fourdspinor$ and $\calz\rightarrow e^{i\alpha}\calz$. Altogether, $\calz$ has weights $(3,1)$ under the dilatation and chiral transformation. By fixing $\vartheta=0$, one is back to the formulation of section \ref{calibrations}. On the other hand, in order to clarify the relation with ordinary four--dimensional Einstein supergravity, a different gauge choice is needed. As in \cite{Kallosh:2000ve}, this is done by isolating  a compensator $\comp$ of  weights $(1,1/3)$ in $\calz$, through the split
\bea\label{split}
\calz=\comp^3\calz^0(z)\ .
\eea
Here $z$ is a set of complex variables parametrizing the deformations of the generalized almost complex structure defined by $\calz^0$. 

Now, the K\"ahler potential $\calk$ is defined by the equation
\bea
\caln=|Y|^2 e^{-\calk/3}\ .
\eea  
The ordinary Einstein supergravity formulation is obtained by imposing \cite{Kugo:1982mr}
\bea\label{eingauge}
\comp=M_{\rm P}\,e^{\calk/6}\ .
\eea
This obviously implies the Einstein frame (\ref{einstcond}), but it also fixes the chiral gauge symmetry. 

We conclude that the K\"ahler potential has the form \cite{effective}
\bea\label{kaehlerpot}
\calk=-3\log\Big(\frac{i\pi}{2}\int_{\calm_6}\langle \calz^0,\bar\calz^0\rangle^{1/3}\langle T,\bar T\rangle^{2/3}\Big)\ .
\eea
On the other hand, the superpotential is given by \cite{effective}
\bea\label{supot}
W =-\pi M^3_{\rm P}\int_{\calm_6}\langle \calz^0(z),F^{(0)}+i\d\calt\rangle\ .
\eea

We can now make contact with the results obtained from the four--dimensional analysis, for which we refer to \cite{Ceresole:2006iq}. First, in the new 4-6 split introduced in (\ref{newsplit}), if one imposes the Weyl--chiral gauge (\ref{eingauge}), the constraint (\ref{algconst}) can be written as
\bea \label{constrEf}
\Im(e^{i\vartheta}\langle \calz^0,F^{(0)}+i\d\calt\rangle)=0\ .
\eea
This implies that $\vartheta$ can be identified with the phase of $W$ (up to a sign ambiguity):
\bea \label{phasesup} 
\Im(e^{i\vartheta} W)=0\quad\Leftrightarrow \quad W=\pm e^{-i\vartheta} |W|\ ,
\eea
in agreement with what is found in \cite{Ceresole:2006iq}. Furthermore, the equation (\ref{adotgen}) for the warping takes the form
\bea \label{adot4d}
\dot A = -\calc=-\frac{1}{M^2_{\rm P}} e^{\calk/2+i\vartheta} W = - \frac{1}{M^2_{\rm P}} e^{\calk/2}|W|\ ,
\eea 
which is also in perfect agreement with \cite{Ceresole:2006iq}. In the last equality of \eqref{adot4d}, we chose the upper sign in \eqref{phasesup}.

Finally, let us note that
from the ten--dimensional flow equations of section \ref{calibrations}, one can show that 
\bea\label{thetadot10d}
\dot\vartheta=\Im \Big(\int_{\calm_6}\langle\del_r{\calz^0}, \delta_{\calz^0}{\calk}\rangle +\int_{\calm_6}\langle\del_r{\calt}, \delta_{\calt}{\calk}\rangle\Big)\ ,
\eea
where, for example, the functional derivative $\delta_{\calz}\calk$ is defined by $\delta\calk=\int_{\calm_6}\langle\delta \calz,  \delta_{\calz} \calk\rangle$. 
The condition (\ref{thetadot10d}) should be compared with the result in four dimensions (see e.g.\ \cite{Ceresole:2006iq})
\bea \label{thetadot}
\dot\vartheta=\Im(\dot{\phi}^i\del_i\calk)\ ,
\eea
where $\phi^i$ are ordinary chiral fields in four dimensions. 

We thus see the emergence of a precise correspondence between the ten--dimensional and the four--dimensional descriptions. 
The ten--dimensional equations \eqref{flow1c}--\eqref{algconst2} also contain flow equations for the fields described by 
$\calz$ and $\calt$, as we elaborate on in appendix \ref{flowdetails}. In the main text we refrain from giving their general form and just apply them (in the next sections) to cases in which one can truncate the four--dimensional spectrum to a finite number of fields. In these cases, the formula of \cite{Ceresole:2006iq}, i.e.\
\bea \label{flowscalars}
\dot{\phi}^i = \frac{1}{M_P^{2}} e^{\calk/2 - i \vartheta} g^{i \bar \jmath}\, \overline{D_j W}\ , 
\eea
is exactly reproduced. 

Before discussing the truncation to a finite number of fields, we would like to make one more comment. The ten--dimensional equation \eqref{flow3c} can be interpreted as a D--flatness condition \cite{effective}. Thus, we see that the closed string D--term is always vanishing for the solutions that we are considering.  


\section{Flow on IIA SU(3)--structure manifolds: general discussion}
\label{truncation}

In this section we apply the general ten--dimensional flow equations in the simplest IIA cases, where the internal manifold $\calm_6$ has SU(3) structure and allows for a consistent finite dimensional parametrization of the polyforms entering the description in ten dimensions. We will show that the general ten--dimensional flow equations can be written as flow equations for a domain wall of the  $\caln=1$ effective theory in four dimensions obtained from (\ref{kaehlerpot}) and (\ref{supot}).


\subsection{General ansatz}
\label{generalsu3}

From now on we assume that the warping function $\warp$, appearing in \eqref{10Dmansatz}, and all the scalar fields evolving in the course of the flow only depend on $r$.

The almost complex structure defined by the SU(3) structure induces a decomposition of the forms $\Lambda^kT^*_{\cal M}=\bigoplus_{p+q=k}\Lambda^{p,q}$. We assume that the internal manifold allows the construction of a certain basis of globally defined forms. Such forms will play the role of the harmonic forms in Calabi--Yau compactifications; their existence was postulated in \cite{glmw,granasup}, and it was recently remarked in \cite{kp,Caviezel:2008ik,ck} that invariant forms on cosets provide a natural set of examples for such a basis.
We denote the forms by the symbols
\bea \label{formbasis}
\nonumber &&\omega_a \in\Gamma(\Lambda^{1,1})\quad\quad a=1,\ldots,n\ ,\\
&&\tilde\omega^a\in\Gamma(\Lambda^{2,2})\quad\quad a=1,\ldots,n\ ,\\
\nonumber&&\alpha_I,\beta^J\in\Gamma(\Lambda^3T^*_{\cal M})\quad\quad I,J=0,\ldots,m\ .
\eea
The basis of even forms is completed by the identity 1 and by a volume form $\d\rm{vol}_0$, which also sets the orientation such that
\bea
\text{Vol}_0\equiv \int\d\text{vol}_0>0\ .
\eea
We assume that one can not build any non-trivial 5-forms from the elements of \eqref{formbasis}, i.e.\ 
\bea \label{omegaalpha}
\omega_a \wedge \alpha_I = 0 = \omega_a \wedge \beta^I\ ,
\eea
and impose the following normalization conditions
\bea
\langle\omega_a,\tilde\omega^b\rangle=\delta_a^b\d\text{vol}_0\ , \qquad \langle\alpha_I,\beta^J\rangle=\delta_I^J\d\text{vol}_0\ .
\eea
We further assume that the basis obeys the following closed differential system 
\begin{equation}
	\begin{split}
		\label{difsystem}
		\d\omega_a= q_{a,I}\beta^I\ ,&\qquad \d\alpha_I=-q_{a,I}\tilde\omega^a\ ,\\
		\d\tilde\omega^a=0\ ,&\qquad \d\beta^I=0\ ,
	\end{split}
\end{equation}
with $q_{a,I}$ which are constant on $\calm_6$.\footnote{Notice that the fact that the constants appearing in the expansion of $\d\omega_a$ and $\d\alpha_I$ are the same is not a restrictive choice but is actually required by the self-consistency of the system (\ref{difsystem}).}

We also define the intersection numbers $\cali_{abc}$ through
\bea
\omega_a\wedge\omega_b\wedge\omega_c=\cali_{abc}\,\d\rm{vol}_0\ .
\eea
In terms of these forms, we consider SU(3) structures given by
\bea\label{su(3)exp}
J=\sum_a e^{2U_a}\omega_a\ , \qquad \Re\Omega=e^\phi(t^I\alpha_I)\ .
\eea
Here $\Re\Omega$ is a generic `stable' form \cite{hitchin_stable} and the complete $(3,0)$-form $\Omega$ can be univocally constructed from it -- see next subsection. The dilaton is introduced 
to get the usual normalization condition
\bea\label{su3normcond}
J\wedge J\wedge J=-\frac{3}{2}\Re\Omega\wedge \Im\Omega\ .
\eea
Finally, we impose that the metric determined by $J$ and $\Omega$ is such that
\bea\label{perpcond}
\langle*_\bindex \alpha_I,\beta^J \rangle=0\ . 
\eea
This implies that $\Im\Omega$ can be expanded in terms of $\beta^I$.

We will see that such a truncation allows to describe flow solutions by an effective four--dimensional action. In section \ref{explicit} we will discuss in detail explicit examples corresponding to compactifications on coset and nearly K\"ahler spaces; let us first, however, discuss the general structure of the flow equations.


\subsection{Ten--dimensional flow from four--dimensional effective theory}

We assume as above that $B$ has only internal components, and we expand
\bea
\NSB=b^a\omega_a\ .
\eea  
Setting $\eta_1=ie^{i(\theta-\vartheta)}\eta_2^*$, where $\vartheta$ is the arbitrary pure gauge phase introduced in section \ref{4Dint}, the NSNS degrees of freedom are contained in
\begin{equation}
	\begin{split}
		\label{NKpurespinors}
		\calz=&\,e^{3\warp-\phi}e^{i( \theta -\vartheta)}e^{iJ +\NSB}=e^{3\warp-\phi}e^{i(\theta -\vartheta)}\exp\big[i(e^{2U_a}-ib^a)\omega_a\big]\ ,\\
		\Re T=&\,e^{-\phi}\Re\Omega=t^I\alpha_I\ .
	\end{split}
\end{equation}
$T$ (and, in particular, its imaginary part $\Im T$) should be considered as a function of its real part $\Re T$, and, hence, of the real parameters $t^I$. To see this \cite{hitchin_stable}, choose a certain coframe $\cof^1,\ldots,\cof^6$. In particular, this allows to define the volume form\footnote{We always use the notation $e^{i\ldots j} = \cof^i \wedge \ldots \wedge \cof^j$.}
\bea \label{volform}
\d\rm{vol}_0 \equiv \cof^{123456}\ .
\eea
Then define
\bea
\hat I_m{}^n=\frac{1}{12}(\Re T)_{k_1k_2k_3}(\Re T)_{k_4k_5m}\epsilon^{k_1\ldots k_5n}\ .
\eea
Now, $\Re T$ is said to be {\em stable} if $\hat I_m{}^n \hat I_n{}^m<0$. In this case,  the almost complex structure defined by $\Re T$ is given by
\bea
I_m{}^n=\frac{\hat I_m{}^n}{\sqrt{-\frac16\, \hat I_k{}^l \hat I_l{}^k}}\ .
\eea
Then, $\Im T$ is given by
\bea
\Im T=-\frac13 I_m{}^n\,\cof^m\wedge \iota_{e_n}\Re T = e^{-\phi} \Im \Omega\ .
\eea
Notice also that, in the above parametrization, the dilaton $e^\phi$ is a function of $U_a$ and $t^I$, and it is determined by (\ref{su3normcond}). In the following we will also need the Hitchin functional
\bea\label{hitfunc}
\calh=\frac{1}{4\text{Vol}_0}\int_{{\cal M}_6} \langle \Re T,\Im T\rangle\equiv \frac{1}{2\text{Vol}_0}\int_{{\cal M}_6} \d{\rm vol}_0 \sqrt{-\frac16\, \hat I_k{}^l \hat I_l{}^k}\ ,
\eea
where the overall factor $\text{Vol}_0^{-1}$ is introduced for later convenience. In our case, $\calh$ should be seen as a function of the parameters $t^I$. Notice that $\calh(t)$ is a homogeneous function of degree 2.

We can expand the RR sector (in the twisted picture) analogously. First we expand both $F^{(0)}$ and $C$  of the decomposition \eqref{splitF} in the basis \eqref{formbasis}. As before, only $C$ varies with $r$. Then we expand
\begin{equation}
	\begin{split}
		\label{eq:fomega}	F^{(0)}=&\,f_0+f^a_2\omega_a+f_{4,a}\tilde\omega^a+f_6\,\d\text{vol}_0\ ,\\
		C=&\,\f4^I\, \alpha_I\ ,
	\end{split}
\end{equation}
with constant $f_0,f^a_2, f_{4,a}$ and $f_6$, or equivalently
\bea
F^\bindex=f_0+f^a_2\omega_a+(f_{4,a}-\f4^I\,q_{a,I})\tilde\omega^a+f_6\,\d\text{vol}_0\ .
\eea

The presence of localized sources like D--branes or orientifolds filling the four--dimen\-sio\-nal space--time would violate the ansatz described in section \ref{generalsu3}, since they would, for example, force the warping and dilaton to be non--constant on $\calm_6$. Smeared sources could solve the problem at the technical level, but their ten--dimensional justification is clearly more problematic. Thus, in this section we do not include 4D space--time--filling localized sources. Then, the Bianchi identities (\ref{splitBIb}) are fulfilled if  
\bea\label{fluxconst}
q_{a,I}f^a_2=0\ .
\eea 
Notice also that some of the $f_{4,a}$ are redundant since the shift
\bea\label{fluxred}
f_{4,a}\rightarrow f_{4,a}+\Lambda^I q_{a,I}
\eea
can be reabsorbed in the constant `axionic' shift $\f4^I\rightarrow \f4^I+\Lambda^I$. 

We can now introduce the following chiral fields of our (superconformal) four--dimensional description
\bea
\rho^a\equiv e^{2U_a}-ib^a\ ,\qquad \tau^I\equiv t^I-i\f4^I\ ,\qquad \comp\equiv e^{\warp-\phi/3}e^{i(\theta-\vartheta)/3}\ .
\eea
The polyforms entering the four--dimensional description are 
\bea
\calz=\comp^3e^{i\rho^a\omega_a}\ ,\qquad \calz^0=e^{i\rho^a\omega_a}\ , \qquad \Re T=(\Re\tau^I)\alpha_I\ , \qquad \calt=\tau^I\alpha_I\ .
\eea 
We can then fix the Einstein frame $\comp=M_{\rm P}e^{\calk/6}$ (and thus $\vartheta=\theta$).
Notice that, comparing our truncation with the one used in \cite{granasup}, we are excluding from the spectrum the axion obtained by dualizing the $B$-field with external legs and possible fluctuations of the RR-potential $C$ and $\Re T$ along $\beta^I$, cf.\ \eqref{NKpurespinors} and \eqref{eq:fomega}.  This truncates the hypermultiplets of \cite{granasup} (see \cite{ck} for the more explicit example of compactifications on coset manifolds) to the chiral fields $\tau^I$ and implies that the description is intrinsically $\caln=1$. 

Defining $V_0\equiv 4\pi\text{Vol}_0$, the superpotential (\ref{supot}) takes the form
\bea\label{cosetsup}
W&=&\frac14\,M_{\rm P}^3V_0\big[f_6-i(f_{4,a}-iq_{a,I} \tau^I)\rho^a-\frac12 f^a_2\cali_{abc}\rho^b\rho^c+\frac{i}{3!}f_0\cali_{abc}\rho^a\rho^b\rho^c\big]\ ,
\eea
while  the K\"ahler potential (\ref{kaehlerpot}) becomes 
\bea\label{cosetkaehler}
\calk&=&-\log\Big[ \cali(\rho+\bar\rho) \Big] - 2\log\Big[ \calh(\tau+\bar\tau)\Big] -3\log V_0\ ,\eea
where
\bea \label{eq:I}
\cali\equiv \frac1{3!}\cali_{abc}(\Re\rho)^a(\Re\rho)^b(\Re\rho)^c
\eea
and $\calh$ is the Hitchin functional (\ref{hitfunc}). Again, we stress that $\calh$ must be seen as a function of $\Re T$ alone, and thus as a function of $t^I=(\Re\tau)^I$. We will see explicit examples of this in the following sections. From (\ref{perpcond}) it also follows that, once one knows $\calh(t)$, $\Im T$ is given by
\bea
\Im T=2\frac{\del\calh}{\del t^I}\,\beta^I\ .
\eea

Using the superpotential \eqref{cosetsup} and the K\"ahler potential \eqref{cosetkaehler} it is possible to show that the flow equations \eqref{flow1c}-\eqref{algconst2} can be expressed as 
\bea \label{dotAtheta}
\dot A=-\frac{1}{ M^2_{\rm P}} e^{\calk/2+i\vartheta} W \ , \qquad \dot\vartheta=\Im(\dot\rho^a\del_a\calk+\dot\tau^I\del_I\calk)\ 
\eea
together with 
\bea
\dot\rho^a & = & \frac{1}{M^2_{\rm P}}e^{\calk/2-i\vartheta}G^{a\bar b} \overline{D_b W}\ , \label{dotrho} \\
\dot\tau^I &=& \frac{1}{ M^2_{\rm P}} e^{\calk/2- i\vartheta}G^{I\bar J} \overline{D_J W}\ , \label{dottau}
\eea
where 
\bea
G_{a\bar b}&\equiv &\frac{\del^2 \calk}{\del \rho^a\del\bar\rho^b}\ , \\
G_{I\bar J}&\equiv &\frac{\del^2 \calk}{\del \tau^I\del\bar\tau^J}\ , \\
D_a W&\equiv &\frac{\del W}{\del \rho^a}+\frac{\del \calk}{\del \rho^a}\,W\ , \\
D_I W&\equiv &\frac{\del W}{\del \tau^I}+\frac{\del \calk}{\del \tau^I}\,W\ .
\eea
The details of the derivation can be found in appendix \ref{flowdetails}.
As anticipated, the equations for the scalars, \eqref{dotrho} and \eqref{dottau}, coincide exactly with the general formula \eqref{flowscalars}. As discussed in 
\cite{Ceresole:2006iq}, the system \eqref{dotAtheta}-\eqref{dottau} can be rewritten
by using the function $\calc$ appearing in \eqref{adot4d}.
With its help the flow equations take the form 
\bea \label{flowc}
\dot A=-\calc \ , \qquad \dot{\phi}^i = 2 g^{i \bar \jmath}\, \partial_{\bar{\jmath}} \calc\ , 
\eea
whereas the equation for $\vartheta$ in \eqref{dotAtheta} is automatically satisfied, cf.\ \cite{Ceresole:2006iq}.

\section{Explicit examples}
\label{explicit}

\subsection{Coset manifolds}
The three possible cosets with non--reducible SU(3) structure are 
\bea
\frac{\text{SU(3)}}{{\rm U(1)}\times {\rm U(1)}}\ , \qquad \frac{\text{Sp(2)}}{{\rm S(U(1)}\times {\rm U(1))}} \ , \qquad\frac{{\rm G_2}}{{\rm SU(3)}}\ .
\eea
Topologically, they correspond to the ``flag manifold'' ${\Bbb F}(1,2;3)$, to $\mathbb{CP}^3$ and to $S^6$, respectively.  AdS$_4$ vacua with these internal spaces have been found in massive IIA in \cite{behr,tomtwistor,klt} and effective theories for those vacua have been described in \cite{kp,ck,paltihouse,Caviezel:2008ik} (some time ago, these spaces were already discussed in the context of string compactifications in \cite{Lust:1986ix}).

Introducing a coframe $\{\cof^1,\ldots, \cof^6\}$ inherited from the parent group, one can construct a volume form as in \eqref{volform} and left--invariant even forms $\omega_a,\tilde\omega^a$, with $a=1,\ldots,b_2+1$, where $b_2$ is the second Betti number (see e.g.\ \cite{klt,ck}). On the other hand, for all these cosets the only left--invariant odd forms are $\alpha,\beta$, defined by
\bea
(\beta+i\alpha)=\frac12\,(\cof^1+i\cof^2)\wedge(\cof^3+i\cof^4)\wedge(\cof^5+i\cof^6)\ .
\eea
The above left--invariant forms define a closed system of the kind (\ref{difsystem}). Thus, for these spaces, $n=b_2+1$ and $m=0$ and one only has a single parameter $\tau$ defined by $\calt=\tau\alpha$. The associated Hitchin functional (\ref{hitfunc}) is given by
\bea
\calh=[(\tau + \bar{\tau})/2]^2\ .
\eea

We now give some details about the three examples. We do not write the form of the structure constants. They can be found in \cite{klt} after some sign adjustments (more precisely, for $\frac{\rm{SU(3)}}{{\rm U(1)}\times {\rm U(1)}}$ and $\frac{{\rm G_2}}{{\rm SU(3)}}$ we switched the sign of $e^1$ and $e^5$ and for $\frac{\rm{Sp(2)}}{{\rm S(U(1)}\times {\rm U(1))}}$ we switched the sign of $e^5$).


\subsubsection{$\frac{\rm{SU(3)}}{{\rm U(1)}\times {\rm U(1)}}={\Bbb F}(1,2;3)$}

In this case, $a=1,2,3$ and the left--invariant metric is given by
\bea
\d s^2_6=e^{2U_1}[(\cof^1)^2+(\cof^2)^2]+e^{2U_2}[(\cof^3)^2+(\cof^4)^2]+e^{2U_3}[(\cof^5)^2+(\cof^6)^2]\ .
\eea
The left--invariant even basis is  given by
\begin{equation}
	\begin{split}
		\omega_1=\cof^{12}\ , \qquad &\omega_2=\cof^{34}\ , \qquad \omega_3=\cof^{56}\ ,\\
		\tilde\omega^1=\cof^{3456}\ , \qquad  &\tilde\omega^2=\cof^{1256}\ , \qquad\tilde\omega^3=\cof^{1234}\ ,
	\end{split}
\end{equation}
with non-vanishing intersection numbers
\bea
\cali_{123}=1\ .
\eea
The constants $q_a$ are given by
\bea
q_a=-1\quad a=1,2,3\ ,
\eea
and the reference volume is
\bea
\text{Vol}_0=2^5\pi^3\ .
\eea


\subsubsection{$\frac{\rm{Sp(2)}}{{\rm S(U(1)}\times {\rm U(1))}}=\mathbb{CP}^3$}

In this case, $a=1,2$ and the left--invariant metric is given by
\bea
\d s^2_6=e^{2U_1}[(\cof^1)^2+(\cof^2)^2+(\cof^3)^2+(\cof^4)^2]+e^{2U_2}[(\cof^5)^2+(\cof^6)^2]\ .
\eea
The left--invariant even basis is  given by
\begin{equation}
	\begin{split}
		\omega_1=\cof^{12}+\cof^{34}\ , \qquad &\omega_2=\cof^{56}\ ,\\
		\tilde\omega^1=\frac12(\cof^{3456}+\cof^{1256})\ , \qquad&\tilde\omega^2=\cof^{1234}\ ,
	\end{split}
\end{equation}
with  non-vanishing intersection numbers
\bea
\cali_{112}=2\ .
\eea
The constants $q_a$ are given by
\bea
q_1=2\ , \qquad q_2=1\ ,
\eea
and the reference volume is
\bea
\text{Vol}_0=2^7\pi^3/3\ .
\eea


\subsubsection{$\frac{{\rm G_2}}{{\rm SU(3)}}=S^6$}
\label{g2coset}

In this case, $a=1$ and the left--invariant metric is given by
\bea
\d s^2_6=e^{2U}[(\cof^1)^2+(\cof^2)^2+(\cof^3)^2+(\cof^4)^2+(\cof^5)^2+(\cof^6)^2]\ .
\eea
The left--invariant even basis is  given by\footnote{The normalization is chosen in order to obtain the same constant $q_1$ as used in section \ref{nk}.}
\begin{equation}
	\begin{split}
		&\omega_1=-\sqrt{3} (\cof^{12}+\cof^{34}+\cof^{56})\ ,\\
		&\tilde\omega^1=-\frac{1}{3 \sqrt{3}}(\cof^{3456}+\cof^{1256}+\cof^{1234})\ ,
	\end{split}
\end{equation}
with  non-vanishing intersection number
\bea
\cali_{111}=-18 \sqrt{3}\ . 
\eea
The constant $q_a$ is given by
\bea
q_1=-6\ ,
\eea
and the reference volume is
\bea
\text{Vol}_0=144\pi^3/5\ .
\eea


\subsection{Application to nearly K\"ahler flows}
\label{nk}


Next we would like to apply the general discussion of section \ref{truncation} to the case where the six--dimensional internal space is nearly \ka. AdS$_4$ vacua with these internal spaces have been found in massive IIA in \cite{behr}; an effective theory for those vacua has been described in \cite{kp}.
More precisely, we assume that $\d s^2_6=e^{2U(r)}\d s^2_0$ for some fixed (i.e.\ $r$-independent) nearly \ka\ metric $\d s^2_0$ (an example would be the space $S^6$ of subsection \ref{g2coset}). Then $J=e^{2U}J_0$ 
and $\Omega= e^{3U} \Omega_0$, with 
\begin{equation}\label{eq:nk0}
	\d J_0 = -3 {\rm Im} \Omega_0 \ ,\qquad 
	\d {\rm Re} \Omega_0= 2 J_0^2\ .
\end{equation}
Moreover, we assume that the fields are only in the singlets of this ${\rm SU}(3)$ structure: 
\begin{equation}
	\NSB = b J_0 \ ,\qquad F^\bindex= f_0 + \f4 J_0^2+\frac16 f_6 J_0^3\ .
\end{equation}
From \eqref{splitBIb}, we see that away from the sources, $f_0$ and $f_6$ should be constant. Since $J_0^2$ is exact (from (\ref{eq:nk0})), we have that $F^\bindex=F^{(0)}+\d C$ with
\bea
F^{(0)}= f_0+\frac16 f_6 J_0^3\ , \qquad C= \frac12 \f4\Re\Omega_0\ .
\eea
Then, we can use as dynamical parameters describing the flow, $A$ and the complex parameters
\bea
\rho\equiv  e^{2U} - i b \ , \qquad \tau \equiv  2 e^{3U-\phi}-i \f4\ .
\eea
Comparing this truncation to the one used in \cite{kp}, we see that we are keeping the complex scalar belonging to the $\caln=2$ vector multiplet, but only two of the four-real scalars forming the $\caln=2$ hypermultiplet used in \cite{kp}. In particular, we are truncating away the hypermultiplet scalars that would be associated to the exact left--invariant 3-form. 

The conformal compensator $\comp$ is given by $\comp^3=e^{3\warp -\phi}e^{i(\theta-\vartheta)}$.
In particular $\calz$, $T$ and $\calt$ are given by
\bea
\calz=\comp^3e^{i \rho J_0}\ , \qquad T = (\Re \tau)\,\Omega_0  \ , \qquad\calt=\tau \,\Re \Omega_0\ .
\eea
Thus, the K\"ahler potential is given by
\bea
\calk&=&-3\log\Big[\frac{\pi i}{2}\int_{\calm_6}\langle e^{i\rho J_0},e^{-i \bar \rho J_0}\rangle^{1/3}\langle T,\bar T\rangle^{2/3}\Big]\cr
&=& -3\log[(\rho+\bar\rho)/2]-4\log [(\tau+\bar \tau)/2]-3\log V_0\  \ ,
\eea
where $V_0=4\pi \text{Vol}_0(\calm_6)$, $\text{Vol}_0(\calm_6)$ being the volume determined by $J_0,\Omega_0$. Imposing the Einstein frame gauge-fixing (\ref{eingauge}) amounts to setting $\vartheta=\theta$ (so that $\comp$ is real) and 
\bea
\warp=\phi-3U-\log(\sqrt{V_0}/M_{\rm P})\ .
\eea
Finally, the superpotential is given by
\bea
W= M^3_{\rm P}V_0\Big(\frac{i}{4} f_0 \rho^3 + \frac32 \rho \tau + \frac14 f_6\Big)\ .
\eea
If we perform the K\"ahler transformation $\calk\rightarrow \calk -3\log V_0$, $W \rightarrow V_0^{3/2} W$ and we  set $M_{P}=\sqrt{V_0}$, we get a somewhat simplified `gauge' choice for the above quantities   
\bea
M_{P}=\sqrt{V_0}\quad \Rightarrow \quad\calk &=& -3\log[(\rho+\bar\rho)/2]-4\log [(\tau +\bar \tau)/2]\ ,\nonumber \\[1ex]
\quad W &=& M^2_{\rm P}\Big(\frac{i}{4} f_0\rho^3+\frac32 \rho \tau +\frac14f_6\Big)\ , \nonumber \\[1ex]
\quad \warp&=&\phi-3U\ \label{Z}. 
\eea

From equations (\ref{dotrho}) and (\ref{dottau}) we get the flow equations for $\rho$ and $\tau$
\bea\label{flowrhot}
\dot\rho=\frac{1}{M^2_{\rm P}}\, e^{\calk/2-i\vartheta}G^{\rho\bar \rho} \overline{D_{\rho} W} \ , \qquad \dot \tau=\frac{1}{M^2_{\rm P}}\, e^{\calk/2-i\vartheta}G^{\tau \bar \tau} \overline{D_{\tau} W}
\eea
and eqs.\ \eqref{dotAtheta} simplify to  
\bea \label{athetadot}
\dot A=-\frac{1}{M^2_{\rm P}} e^{\calk/2+i\vartheta} W \ , \qquad \dot\vartheta=\Im(\dot\rho\partial_\rho\calk+\dot \tau \partial_\tau \calk)\ .
\eea

The flow equations have the following AdS$_4$ solutions \cite{behr} for arbitrary negative values of $f_0$ and positive values of $f_6$: 
\begin{equation} \label{bgrgeneral}
	\begin{array}{c}\vspace{.3cm}
		b e^{-2U}= - \frac 1 {\sqrt{15}} \ ,\qquad \vartheta= 
		{\rm arcsin}\left(\frac14\right) \ , \\ 
		\f4 = - \left(\left(\frac{f_6}{5}\right)^{2/3} \left(\frac{-f_0}{2}\right)^{1/3} \right)
		\ ,\\
		 \phi=\frac16 \ln\left(\frac{15^{5/2}}{6 \, f_6\,  (-f_0)^5} \right) \ ,\qquad U=\ln\left( \left(-\frac{f_6}{f_0}\right)^{1/6} 3^{1/4} 5^{1/12} 2^{-5/6}\right)\ .	
	\end{array}	
\end{equation}

It is also straightforward to check that the usual D2-brane metric
\begin{equation} \label{D2brane}
	\d s^2 = H_2^{-1/2}\d s^2_{{\rm Mink}_{2,1}}+ 
	H_2^{1/2}(\d q^2 + q^2 \d s^2_0)\ ,\qquad H_2 = 1 + \frac Q{q^5}\ ,
\end{equation} 
solves the flow equations, \eqref{flowrhot} and \eqref{athetadot}. Comparing \eqref{D2brane} with \eqref{10Dmansatz} (using $ds^2_6= e^{2U} ds_0^2$ and $\warp=\phi-3U$), we get
\begin{equation}\label{eq:D2}
	e^\phi=H_2^{1/4} \ ,\qquad e^{A}=H_2^{1/4} q^3 \ ,\qquad e^{U}= q H_2^{1/4}\ ,\qquad
	dr = dq H_2^{3/4} q^3\ .
\end{equation}
This is a solution in case $\vartheta = \pi$, $f_0 = \f4=b=0$ and one makes the identification $Q=\frac15 f_6$.


\section{Inclusion of D--branes}
\label{inclusion}

As we already mentioned in the introduction, we allow in this paper for the presence of D--brane sources. Our analysis in section 
\ref{calibrations} was performed directly in ten dimensions. As such, it would permit any kind of brane sources. However, we will limit our discussion here to branes that can be introduced consistently with the truncations we considered in 
sections \ref{truncation} 
and \ref{explicit}. In order not to excite any KK modes, the branes will then have to sit at some fixed value of $r$ and will have
to be either D8--branes, which of course wrap the whole internal space ${\cal M}_6$, or (perhaps less rigorously) lower--dimensional branes, smeared appropriately along the internal directions. In terms of the notation in section \ref{calibrations}, we will hence take $j=0$ and keep only $j_r \neq 0$. 

We will now discuss how the flows are modified by the presence of these D--branes. We will first explain this from the point of view of the bulk supersymmetry equations, namely the flow equations (\ref{flow1c})--(\ref{algconst2}). We will then compare this with the conditions coming from supersymmetry on the D--brane itself. As we will see, one of the resulting conditions always follows automatically from the conditions in the bulk; in the SU(3) structure case, all the conditions arising from calibrating the branes actually follow from the bulk conditions.

From the point of view of the flow equations, the presence of these sources has to be taken into account by imposing that the fields jump in the appropriate way. If our solutions were non--supersymmetric, we would have to impose, for example, the Israel junction conditions on the metric \cite{Israel:1966rt}. Once supersymmetry is imposed, however, the equations of motion for the metric and the dilaton follow from the equations of motion and Bianchi identities for the fluxes \cite{Gauntlett:2002fz,lt,Gauntlett:2005ww}. Hence, the conditions on the first derivatives of the metric will automatically follow from the jump conditions required for supersymmetry and from the flux equations. In fact, the supersymmetry conditions are just valid everywhere, even at the locus where the brane is present. So they do not give rise to any jump. The equations of motion for the flux, on the contrary, do contain delta--like sources localized on the brane.  
Since we are considering only branes that do not break the internal symmetries, from (\ref{splitBIb}) and (\ref{eq:frc}) we get 
\bea\label{backBI}
\del_r F^{(0)}=-j^\bindex_r\ .
\eea
Since the branes are localized in the $r$ direction, we can simply take $j^B_r = - \delta(r-r_0) \Delta F^{(0)}$, which introduces a jump 
\begin{equation}
	\Delta F^{(0)}
\end{equation}
in the background flux $F^{(0)}$. This changes $W$. Hence, the entire flow behaves differently after having crossed the brane.

Now we notice that the constraint \eqref{constrEf} should be valid on both sides of the brane, implying that the brane position $r_0$ should be such that 
\begin{equation}\label{eq:braneconst}
	{\rm Im} (e^{i\vartheta}\langle\calz^0, \Delta F^{(0)}\rangle)= 0 \ .
\end{equation}

So far we have looked at the conditions on the brane that follow from the bulk supersymmetry equations, namely the flow equations (\ref{flow1c})--(\ref{algconst2}). The total string theory action contains both the bulk and also brane terms. A priori, supersymmetry on the brane has to be imposed separately. We will see, however, that one of the resulting conditions is just equivalent to (\ref{eq:braneconst}).

Remember that a supersymmetric D--brane at a radius $r_0$ and wrapping $(\Sigma,F^{\rm (wv)})$ in $\calm_6$ should satisfy  \cite{gencal,branesuppot} 
\bea\label{calcond}
[\Re(e^{i\vartheta}\calz)|_\Sigma\wedge e^{F^{\rm (wv)}}]_{\rm top}=e^{3\warp-\phi}\sqrt{\det((g+B)|_\Sigma+F^{\rm (wv)})}\ ,
\eea
or equivalently
\bea\label{DWcalcond}
&&[(\mathbb{X}\cdot T)|_\Sigma\wedge e^{F^{\rm (wv)}}]_{\rm top}=0\ 
\ ,\qquad \forall \mathbb{X}\in T_{\cal M}\oplus T^*_{\cal M} \ ,\\
\label{DWcalcond2}
&&[\Im(e^{i\vartheta}\calz)|_\Sigma\wedge e^{F^{\rm (wv)}}]_{\rm top}=0\ .
\eea
Suppose now that the cycle $(\Sigma, F^{\rm (wv)})$ is dual to the current $j^B_r= - \delta(r-r_0) \Delta F^{(0)}$. Equation (\ref{DWcalcond2}) can then be rewritten as 
\bea
[\Im(e^{i\vartheta}\calz)|_\Sigma\wedge e^{F^{\rm (wv)}}]_{\rm top}\sim  \Im(e^{i\vartheta}\langle\calz,j^\bindex_r\rangle) \sim \Im(e^{i\vartheta}\langle\calz, \Delta F^{(0)} \rangle)\ . 
\eea
Hence, imposing (\ref{DWcalcond2}) is the same as imposing (\ref{eq:braneconst}). 

As for the condition in (\ref{DWcalcond}), it is not clear to us whether it follows in full generality from the flow equations (\ref{flow1c})--(\ref{algconst2}). It does follow for the setup in section \ref{truncation} though. To see this, notice that for SU(3) structures (\ref{DWcalcond}) reads (at $r=r_0$)
\begin{equation} \label{su3cal}
	[\iota_X \Omega \wedge \Delta F^{(0)}]_6=0 \quad , \quad [\Omega \wedge \Delta F^{(0)}]_5 = 0 \ ,
\end{equation}
for $X \in T_{\cal M}$. The validity of \eqref{su3cal} follows from (\ref{eq:fomega}) and from the fact that 
$\alpha_I \wedge \omega_a = 0 $, cf.\ \eqref{omegaalpha}. 


\subsection{The D--brane modified c-theorem}
\label{ctheorem}

Having introduced explicit brane sources in our setup, we want to check now that the c--theorem \cite{Girardello:1998pd,FGPW} is still valid, as one expects to be the case.
From the analysis of \cite{FGPW}, we know that,  for any domain-wall (DW) solution of four--dimensional gravity coupled to `reasonable' matter,  $\dot A$ is a monotonically decreasing function of the radial direction. In our case, this is equivalent to saying that the function $\calc$ introduced in \eqref{adotgen}
is a monotonically increasing function. In the truncated theories, one can easily see this in the absence of D--branes by taking its radial derivative and using the second equation of (\ref{flowc}), following \cite{Ceresole:2006iq}. In this way, one  obtains 
\bea\label{ct}
\dot \calc=4G^{i\bar\jmath}\partial_i \calc\bar\del_{\bar\jmath} \calc\geq 0\quad\quad\text{(in absence of D--branes)}\ .
\eea

We now want to see if and how this result changes in the presence of supersymmetric D--brane DW's localized at some radial position $r_0$.  Notice that such a D--brane actually causes a jump in the flux quanta and then changes the four--dimensional effective theory on the two sides. Thus the setting is intrinsically ten-dimensional and, a priori, one cannot apply the above arguments about the monotonicity of $\calc$ to this case. Nevertheless, we have a fully ten--dimensional description of the flow and of the four--dimensional objects entering the definition of $\calc$ and then we can directly compute $\dot \calc$ using it.  Indeed, assuming that a standard four--dimensional description of the flow is still possible, we see that (\ref{ct}) is modified in the following way
\bea\label{ct2}
\dot \calc=\frac{T_{\rm D}}{2M^2_{\rm P}}\,\delta(r-r_0)+ 4G^{i\bar\jmath}\partial_i \calc\bar\del_{\bar\jmath} \calc\ ,
\eea
where the D--brane contribution is proportional to its tension $T_{\rm D}$. This, in turn, can be inferred from (\ref{supot}), (\ref{backBI}) and (\ref{calcond}) to be positive and given by
\begin{equation}
	\label{dtens}
	T_{\rm D}=2\pi M^3_{\rm P}\int \d r\int_{\calm_6}\langle\Re(e^{\calk/2+i\vartheta}\calz^0), j^\bindex_r\rangle=2\pi M^3_{\rm P}\int_{\Sigma}\Re(e^{\calk/2+i\vartheta}\calz^0)|_\Sigma\wedge e^{F^{\rm (wv)}}\geq 0\ . 
\end{equation}
Thus, we still have
\bea\label{ct3}
\dot \calc\geq 0\ ,
\eea
or, better, $\calc$ jumps by $T_{\rm D}/(2M^2_{\rm P})$ when it hits the D--brane.

Using the formula (\ref{ct2}), one can see that the total tension of the full supergravity+D--brane DW configuration can be expressed as
\bea\label{generaltension}
T_{\rm DW}=2M^2_{\rm P}\Delta \calc=2M^2_{\rm P}(\calc_{+\infty}-\calc_{-\infty})\ ,
\eea
which is always positive because of (\ref{ct3}). As compared to the tension obtained purely in supergravity \cite{Ceresole:2006iq}, \eqref{generaltension} additionally contains the explicit contribution from the branes \eqref{dtens}. More explicitly, 
\bea
T_{\rm DW}&=& 2M^2_{\rm P} \int_{-\infty}^{\infty}\d r\,\dot \calc = 2M^2_{\rm P}\lim_{\varepsilon\rightarrow 0} \Big[(\calc_{+\infty}-\calc_{r_0+\varepsilon})+ (\calc_{r_0-\varepsilon}-\calc_{-\infty})+\int_{r_0-\varepsilon}^{r_0+\varepsilon}\d r\,\dot \calc \Big] \nonumber \\[2ex]
&=&  2M^2_{\rm P}\lim_{\varepsilon\rightarrow 0}\Big[(\calc_{+\infty}-\calc_{r_0+\varepsilon})+ (\calc_{r_0-\varepsilon}-\calc_{-\infty})\Big]+T_{\rm D}\ ,
\eea
where the first term corresponds to the supergravity contribution.


\begin{acknowledgments}
We would like to thank M.~Berg, D.~Cassani, G.~Dall'Agata, F.~Denef, P.~Koerber, M.~Petrini, D.~Tsimpis for useful discussions and M.~Ammon for collaboration at an early stage. L.M.~thanks P.~Koerber, P.~Smyth and S.~Vaul\`a for early collaboration on the
present topic and private communications.
This work is supported in part by the Excellence Cluster ``The Origin and the Structure of the Universe'' in Munich. M.H.~is supported by the German
Research Foundation (DFG) within the Emmy--Noether--Program (grant number:
HA 3448/3-1). A.~T.~is supported in part by DOE grant DE-FG02-91ER4064.
\end{acknowledgments}



\begin{appendix}

\section{Spinors, differential forms, generalized complex geometry} 
\label{notconv}

A detailed description of the conventions, supersymmetry transformations and equations of motion that we use in this paper can be found e.g.\ in the appendix of \cite{Lust:2008zd}. 

Let us start with a few details about how to deal with spinors in
the various dimensions we are interested in. We start from the decomposition of ten--dimensional spinors in a spacetime of the form ${\Bbb R}^{1,2}\times {\cal M}_7$. The spinor representation in three and seven dimensions has dimension 2 and 8 respectively; hence, the usual  tensor product of gamma matrices will not work. One (standard) way to proceed is to introduce an auxiliary two--dimensional space and write
the gamma matrices as  
\begin{equation}
	\begin{split}\label{eq:gamma37}
		\Gamma_{\ul\mu} &=  \gamma_{\ul\mu} \otimes \sigma_3\otimes \bbone\ ,\\
		\Gamma_{\ul{m}} &=  \bbone \otimes \sigma_1 \otimes \hat\gamma_{\ul{m}}\ ,
	\end{split}
\end{equation} 
where $\sigma_i$ are Pauli matrices acting on the auxiliary two--dimensional space, $\gamma_{\ul\mu}$ is a basis of (real) gamma matrices in three dimensions, and $\hat\gamma_{\ul{m}}$ a basis of (purely imaginary) gamma matrices in seven dimensions (underlining the indices identifies the corresponding coordinates as the flat ones). In the basis (\ref{eq:gamma37}), a ten--dimensional Majorana spinor $\epsilon$ is such that $\epsilon^*= B \epsilon$, where
$B= \bbone \otimes \sigma_3 \otimes \bbone$. The chirality operator is given by $\Gamma= 
- \bbone \otimes \sigma_2 \otimes \bbone$. The two ten--dimensional supersymmetry parameters, which are Majorana--Weyl, can then be written as
\begin{equation}\label{eq:eps}
	\epsilon_1 = \xi_1 \otimes {1 \choose -i} \otimes \chi_1
	\ ,\qquad
	\epsilon_2 = \xi_2 \otimes {1 \choose\pm i} \otimes \chi_2\ ,
\end{equation}
where $\pm$ refers to IIA/IIB, respectively. Here, $\xi_{1,2}$ are two three--dimensional spinors on ${\Bbb R}^{1,2}$, which, for an ${\cal N}=1$ solution, should be taken to be equal, $\xi_1 = \xi_2$. Moreover, $\chi_{1,2}$ are two seven--dimensional Majorana spinors. A slight simplification in section \ref{calibrations} is that we have taken these spinors to have equal norm:
\begin{equation}\label{eq:chi2}
	 \chi_1^\dagger \chi_1 = \chi_2^\dagger \chi_2 
	\equiv |\chi|^2 \ .
\end{equation}
This condition is needed in order to have calibrated branes in a background \cite{branesuppot}. For AdS$_4$ vacua, it can be shown directly from the equations of motion \cite{scan}.

In the main text, we have further split ${\cal M}_7$ as a foliation with generic leaves ${\cal M}_6$, parameterized by $r \in {\Bbb R}$. We then need to split the spinor representation further. The basis we used so far was convenient for the 3+7 split we considered at the beginning of section \ref{calibrations}; in particular, it made it possible to write the $\Psi$ in such a compact form as (\ref{eq:g2g2}).
In order to clarify the relation with the split into 4+6 dimensions, however, it is convenient to use a different ten--dimensional spinorial representation. In this new basis, the ten--dimensional gamma matrices have the following $3+1+6$ dimensional split  
\bea
\Gamma_{\ul\mu}&=&\gamma_{\ul\mu}\otimes\sigma_1\otimes\bbone\ ,\cr
\Gamma_{\ul{r}}&=&\bbone\otimes\sigma_3\otimes\bbone\ ,\cr
\Gamma_{\ul{m}}&=&\bbone\otimes\sigma_2\otimes\hat\gamma_{\ul{m}}\ ,
\eea
where $\gamma_{\ul\mu}$ are three--dimensional real gamma matrices and $\hat\gamma_{\ul{m}}$ are now six--dimensional imaginary gamma matrices. In this basis, the ten--dimensional chirality matrix is $\Gamma=\bbone\otimes \sigma_2\otimes \hat\gamma$, where $\hat\gamma$ is the six--dimensional chirality operator on $\calm_6$.  The MW Killing spinors $\epsilon_{1,2}$ are real and have the form
\bea
\epsilon_{1,2}=\xi\otimes\Big[\frac12\left(\begin{array}{c} 1 \\ i \end{array} \right)\otimes \eta_{1,2}+ \text{c.c.\ }\Big]\ ,
\eea
where $\eta_{1,2}$ are chiral spinors in six dimensions, such that $\hat\gamma\eta_{1}=\eta_{1}$ and $\hat\gamma\eta_{2}=\mp\eta_{2}$ in IIA/IIB. We can also write $\epsilon_{1,2}=\fourdspinor \otimes\eta_{1,2}+$c.c., where the four-dimensional chiral spinor $\fourdspinor$ satisfies the projection condition given in (\ref{firstproj}).

Let us now recall some basic aspects of the formalism of 
generalized geometry -- for more detailed discussions see e.g.~\cite{gualtieri} and \cite[Sec.~3]{scan}. First, the basic objects of this formalism are {\em polyforms}, i.e.\ formal sums of forms of different degree. One can then define the unipotent operator $\lambda$ acting on them as follows
\bea \label{eq:lambda}
\lambda(\d y^{m_1}\wedge \ldots\wedge \d y^{m_k})=\d y^{m_k}\wedge \ldots\wedge \d y^{m_1}\ .
\eea
This can be used to define  a natural pairing (often called ``Mukai pairing'') between two polyforms $\alpha$ and $\beta$:
\bea\label{mukai}
\langle\alpha,\beta\rangle\equiv [\alpha\wedge\lambda(\beta)]_{\rm top}\ ,
\eea  
where, on an $n$-dimensional space, $[\ldots]_{\rm top}$ selects the form of degree $n$.
Thus, the Mukai pairing maps a pair of polyforms to a density. In seven dimensions it is symmetric, while in six dimensions it is antisymmetric. 

The polyforms can be seen as spinors of the generalized tangent bundle $T_{\cal M}\oplus T^*_{\cal M}$. We denote with $\mathbb{X}$ the generic element (or section) of $T_{\cal M}\oplus T^*_{\cal M}$. Writing more explicitly $\mathbb{X}=X+\xi$, with $X\in T_{\cal M}$ and $\xi\in T^*_{\cal M}$, the Clifford action of $\mathbb{X}$ on a polyform $\alpha$ is given by
\bea
\mathbb{X}\cdot\alpha= \iota_X\alpha+\xi\wedge \alpha\ .
\eea
In six dimensions, the complex polyforms $\psone$ and $\pstwo$, or their rescaled and twisted redefinition $T$ and $\calz$, used in the paper are special because they are {\em pure}, i.e.\ they are annihilated by six-dimensional subspaces of $(T_{\cal M}\oplus T^*_{\cal M})\otimes\mathbb{C}$, $L_1$ and $L_2$, respectively. Each of them defines a generalized almost complex structure \cite{hitchin,gualtieri}, which can be used to define a decomposition of the space of polyforms \cite{gualtieri}. For example, we use $\calz$ to define the following decomposition
\bea \label{gdec}
\bigoplus_{p} \Lambda^p T_{\cal M}^*\otimes \mathbb{C}=\bigoplus_{k}U_k\ ,
\eea
where 
\begin{equation}\label{eq:uk}
	U_{3-k}=\Lambda^k \bar{L}_2 \cdot \calz\ . 
\end{equation}
By construction $\calz$ and $T$  define an SU(3)$\times$SU(3) structure, which is equivalent to requiring that
\bea
T\in U_0\ 
\eea
(recall that $\calz$ is in $U_3$ by the definition (\ref{eq:uk})). 

$T$ can also be used to introduce a decomposition similar to (\ref{eq:uk}): 
\begin{equation}\label{eq:vk}
	V_{3-k}=\Lambda^k \bar{L}_1\cdot T\ .
\end{equation}
When $T$ and $\calz$ together define an SU(3)$\times$SU(3) structure, one can refine the two decompositions (\ref{eq:uk}) and (\ref{eq:vk})
by taking their intersection $U_{k,r}= U_k \cap V_r$. One gets a 
``generalized Hodge diamond''
\bea\label{ghodge}
\begin{array}{ccccccc}
&&&  U_{0,3}  &&& \\
&&U_{1,2}&&U_{-1,2}&&\\
&U_{2,1}&&U_{0,1}&&U_{-2,1}&\\
U_{3,0}&&U_{1,0}&&U_{-1,0}&&U_{-3,0}\\
&U_{2,-1}&&U_{0,-1}&&U_{-2,-1}&\\ 
&&U_{1,-2}&&U_{-1,-2}&&\\
&&&  U_{0,-3}  &&&
\end{array}\quad .
\eea
This is not quite the usual Hodge diamond, in spite of its shape. Its elements are in general not forms of a single degree, as in that case. The peculiar degrees are a consequence of the conventions chosen in \eqref{eq:uk}. In spite of these peculiarities, this basis is useful in the computations presented in the next appendix, because of some nice properties that it enjoys, as we now explain. 

Remember that $\calz$ and  $T$ contain the complete information about metric and $B$-field (as well as dilaton and warping). By introducing the twisted Hodge--star operator
\bea\label{twistedhodgestar}
*_\bindex\equiv e^{\bindex}*\lambda e^{-\bindex}\quad\quad (\Rightarrow *^2_\bindex=-1)\ ,
\eea
the decomposition  (\ref{ghodge}) has the property
\bea
*_\bindex \alpha_{k,r}=i(-)^{(k+r+1)/2}\alpha_{k,r}\quad\quad \forall \alpha_{k,r}\in U_{k,r}\ .
\eea 
Another useful property is the following:
\bea
\langle *_\bindex\alpha,\beta\rangle=(e^{-\bindex}\alpha)\cdot(e^{-\bindex}\beta)\d\text{vol}_6\ ,
\eea
where $\d\text{vol}_6$ is the canonical volume form $\sqrt{g}\d^6 y$ defined by the metric and
\bea
(e^{-\bindex}\alpha)\cdot(e^{-\bindex}\beta)=\sum_k\frac{1}{k!}(e^{-\bindex}\alpha)_{m_1\ldots m_k}(e^{-\bindex}\beta)^{m_1\ldots m_k}\ .
\eea

Finally, let us recall the explicit form of the pure spinors (\ref{6dpurespin}) in the SU(3) structure case. A more general explicit form, valid for the generic SU(3)$\times$SU(3) structure case, can be found e.g.~in \cite{mpz} (up to adapting the conventions). SU(3) structure means that the internal spinors are proportional: $\eta_1=ie^{i\theta}\eta^*_2$ for IIA and 
$\eta_1=ie^{i\theta}\eta_2$ for IIB, for some (possibly point--dependent) phase $e^{i\theta}$. 
We can then introduce the normalized spinor $\eta=\eta_1/|\eta_1|$ and use it to construct the following tensors on $\calm_6$
\be\label{jomega}
J_{mn}\, =\, i\eta^\dagger\gamma_{mn}\eta\quad\quad\quad\quad\Omega_{mnp}\, =\, \eta^T\gamma_{mnp}\eta\ ,
\ee
which satisfy 
\begin{equation}
 	 \Omega \wedge J = 0 \ ,\qquad 
   \frac16 J\wedge J\wedge J\, =\, - \frac i8\Omega\wedge\bar\Omega\, =\, \d\text{vol}_6\ .
\end{equation}
$J$ is the two-form associated with the almost complex structure $J^m{}_n$, with respect to which $\Omega$ is a $(3,0)$-form. Together $J$ and $\Omega$ provide an alternative definition of the SU(3) structure of the configuration. 

In this case, the pure spinors $\Phi_1$ and $\Phi_2$ take the form
\begin{equation}
	\begin{split}
		\label{su3ps}
		\Phi_1\, =\, e^{3\warp-\phi-i\theta}\Omega\qquad & \qquad \Phi_2\, =\,e^{3\warp-\phi+i\theta}e^{iJ}\qquad\, \text{in IIA}\ ,\\
		\Phi_1\, =\, e^{3\warp-\phi+i\theta}e^{iJ}\qquad& \qquad \Phi_2\,=\,e^{3\warp-\phi-i\theta}\Omega\qquad\ \ \text{in IIB}\ .
	\end{split}
\end{equation}
Notice that often one is not interested in the overall phase of $\Omega$. Then the factor $e^{-i\theta}$ can be absorbed by a redefinition of $\Omega$, as we did in section \ref{truncation}. 
%


\section{Comparison between ten--dimensional and four--dimensional flow equations}
\label{flowdetails}

In this appendix we discuss in more detail the relation between the ten--dimensional flow equations derived in section \ref{calibrations} and their interpretation from a four--dimensional point of view. We first discuss this aspect in full generality in section \ref{flowdetails1} and then we specialize to the truncated IIA theories of section \ref{truncation} in section \ref{flowdetailscosets}.

\subsection{Ten--dimensional flow equations in four--dimensional form}
\label{flowdetails1}

Using the polyforms introduced in (\ref{calz}), the flow equations (\ref{flow1c})-(\ref{flow4c}) take the form
\bea
\d(e^{4\warp}\Re T)&=&e^{4\warp}*_\bindex F+\partial_r(\Re\calz)
+3(\partial_r A)\Re \calz \ ,\label{flow1d}\\
\d(\Re\calz)&=&-e^{2\warp}*_\bindex F^\bindex_r\ ,\label{flow2d}\\
\d(e^{2\warp}\Im T)&=&0\ ,\label{flow3d}\\
\d(\Im\calz)&=&\partial_r(e^{2\warp}\Im T)+2 (e^{2\warp}\Im T) \partial_r A \label{flow4d}\ ,
\eea
which must be supplemented by the algebraic condition (\ref{algconst2}).

For notational simplicity, in this section we will work with $\calz$ as introduced in (\ref{calz}), without introducing the redundant phase $\vartheta$ as in (\ref{znew}). Furthermore, in order to simplify the form of the following equations, let us introduce the {\em densities}
\bea\label{nw}
N=\frac{i\pi}{2}\langle \calz,\bar\calz  \rangle^{1/3}\langle T,\bar T\rangle^{2/3}\ , \qquad W_{\rm sc}=-\pi\langle \calz,F+i\d\Re T\rangle\ ,
\eea
where the subscript ``sc'' indicates that this is the density of the superconformal superpotential.
From the flow equations (\ref{flow1d})-(\ref{flow4d}), after having imposed the Einstein frame condition (\ref{einstcond}), it is possible to show that 
\bea
(\del_r\calz)_1&=&-2 e^{4\warp}*_\bindex (F+i\d \Re T)^*_1\ ,\label{holflow3}\\
(\del_r\calt)_0&=& -ie^{-2\warp}*_\bindex(\d\bar\calz)_0 -2\Re T\, \frac{W_{\rm sc}}{N}\ ,\label{holflow1}\\
(\del_r\calt)_{-2}&=&0\ ,\label{holflow2}
\eea
where the subscripts ${}_k$ refer to the decomposition introduced in (\ref{gdec})\footnote{In this appendix we use star and bar interchangeably for complex conjugation.}, as well as
\bea
&&\dot A+\frac{W_{\rm sc}}{N}+\frac{\del_r N}{2N}=0\ ,\label{adotgenn}\\
&& e^{-4\warp}\Im(i\langle\del_r\calz,\bar \calz \rangle)+4e^{2\warp}\langle\Im T, F^\bindex_r \rangle=0\label{dthetagen}\ .
\eea
Conversely, if we supplement the set of equations (\ref{holflow3})--(\ref{dthetagen}) by (\ref{flow3d}) and by (\ref{algconst2}), which can also be rewritten as
\bea\label{algconstbis}
\Im W_{\rm sc}=0\ ,
\eea
we can reconstruct (\ref{flow1d})--(\ref{flow4d}).

Let us now discuss the four--dimensional interpretation of the ten--dimensional equations  (\ref{flow3d}) and (\ref{holflow3})-(\ref{algconstbis}). First, as explained in \cite{effective} (see also \cite{casbil}), (\ref{flow3d}) has a clear interpretation as D--flatness condition associated to the gauging of the full tower of KK-modes which are charged under the RR-gauge transformations. Then, in section \ref{4Dint} we already mentioned the relation between (\ref{algconst}) --- and thus (\ref{algconstbis}) --- and the four--dimensional equation (\ref{phasesup}). In the same way, by integrating (\ref{dthetagen}), using the new $\calz$ introduced in (\ref{znew}) and isolating and fixing the compensator $Y$ as in (\ref{split}) and (\ref{eingauge}), one gets (\ref{thetadot10d}) which, as already discussed, is directly related to the four--dimensional equation (\ref{thetadot}). Moreover, the equation \eqref{adotgen} for the warp factor can be reproduced by integrating \eqref{adotgenn}.

On the other hand, we would like to interpret (\ref{holflow3}) and (\ref{holflow1}) as flow equations of the kind (\ref{flowscalars}) for the `chiral fields' $\calz$ and $\calt$. A step in this direction can be made introducing the formal quantity 
\bea
\calk_{\rm sc}=-3\log N\ .
\eea
Then, one can write (\ref{holflow3}) and (\ref{holflow1}) as follows
\bea
(\del_r\calz)_1&=&\frac{2}{\pi}e^{4\warp} *_\bindex(\cald_{\calz_1}W_{\rm sc})^*\ ,\label{holflow3b}\\
(\del_r\calt)_0&=&  \frac{1}{\pi}e^{-2\warp} *_\bindex(\cald_{\calt_0}W_{\rm sc})^*\ ,\label{holflow1b}
\eea
where now
\bea
\cald_{\calt_0,\calz_1}W_{\rm sc}&\equiv &\delta_{\calt_0,\calz_1}W_{\rm sc}+W_{\rm sc}\delta_{\calt_0,\calz_1}\calk_{\rm sc}\ ,
\eea
with, for example, $\delta_{\calt_0}W_{\rm sc}$ defined by $\delta W_{\rm sc}=\langle\delta\calt_0,\delta_{\calt_0}W_{\rm sc}\rangle$. 

The equations (\ref{holflow3b}) and (\ref{holflow1b}) clearly resemble the four--dimensional flow equations (\ref{flowscalars}), although they are obviously not exactly of the same form.\footnote{On the other hand, (\ref{holflow2}) does not seem to have an obvious four--dimensional analog and should be seen as an additional consistency condition.} This is mainly because we are considering an untruncated theory which is still intrinsically ten--dimensional. On the other hand, the analogy between (\ref{holflow3b})-(\ref{holflow1b}) and (\ref{flowscalars}) is also not accidental and, indeed, in the next subsection we will see how in the truncated theory of section \ref{truncation}, they exactly reproduce the expected four--dimensional equations  (\ref{flowscalars}).

\subsection{Details of the derivation of the four--dimensional flow equations}
\label{flowdetailscosets}

We give some details of the derivation of the flow equations \eqref{dotrho} and \eqref{dottau} from the ten--dimensional equations (\ref{flow1c})-(\ref{algconst2}). It is convenient though to use the alternative formulation of  (\ref{flow1c})-(\ref{algconst2}) given in (\ref{holflow3})-(\ref{algconstbis}), supplemented by (\ref{flow3d}), keeping in mind that they should be expressed in terms of the new  $\calz$ defined in (\ref{znew}), which practically corresponds to rewriting them by simply substituting $\calz$ with $e^{i\vartheta}\calz$.
 
First of all let us observe that, since $\d \beta^I=0$, it is easy to see that the D--flatness condition (\ref{flow3d}) is automatically satisfied. Furthermore, given the truncation introduced in section \ref{truncation}, (\ref{holflow2}) is identically satisfied and  (\ref{adotgenn})-(\ref{algconstbis}) just boil down to (\ref{adot4d}), (\ref{thetadot}) and (\ref{phasesup}). Thus, the only ten--dimensional equations that remain to be discussed are (\ref{holflow3}) and (\ref{holflow1}).

We start by considering (\ref{holflow3}), which we expand in elements of $U_1$. First, notice that a basis of those elements of $U_3\oplus U_1$, which are needed for our problem, is given by\footnote{We hope that using these names does not lead to confusion with the spinors we are using in the main text and appendix \ref{notconv}.}
\bea\label{basis1}
\chi\equiv e^{i\rho^b\omega_b}\ , \qquad \psi_a\equiv \omega_a\wedge e^{i\rho^b\omega_b}=-i\frac{\del}{\del\rho^a}\chi\ .
\eea
We can write the following non--vanishing Mukai pairings of its elements 
\bea
\langle\chi,\bar\chi\rangle=-8i\cali\,\d\text{vol}_0\ , \qquad \langle\psi_a,\bar\chi\rangle=-4\cali_{a}\,\d\text{vol}_0\ , \qquad\langle\psi_a,\bar\psi_b\rangle=-2i\cali_{ab}\,\d\text{vol}_0\ ,
\eea
where ${\cal I}$ was defined in (\ref{eq:I}), and we introduced 
\bea
\cali_a&\equiv &2\frac{\del\cali}{\partial\rho^a}=\frac12\cali_{abc}(\Re \rho)^b(\Re\rho)^c\ ,\cr
 \cali_{ab}&\equiv &4\,\frac{\del^2\cali}{\partial\rho^a\partial\rho^b}=\cali_{abc}(\Re \rho)^c\ .
\eea
We can then replace the elements $\psi_a$ with
\bea
\sigma_a=\psi_a+\frac{i}{2}\,\frac{\cali_a}{\cali}\chi
\eea
and obtain the new Mukai pairings
\bea
\langle\chi,\bar\chi\rangle=-8i\cali\,\d\text{vol}_0\ , \qquad \langle\sigma_a,\bar\chi\rangle= 0\ , \qquad\langle\sigma_a,\bar\sigma_b\rangle=-2i\calg_{ab}\d\text{vol}_0\ ,
\eea
where
\bea
\calg_{ab}\equiv \cali_{ab}-\frac{\cali_a\cali_b}{\cali}\ .
\eea
We now use the elements $\sigma_a$ to expand the complex conjugate of (\ref{holflow3}). On the one hand, it is easy to see that
\bea
(\partial_r\bar\calz)_{-1}=Y^3[-i(\dot\rho^a)^*\omega_a\wedge \bar\chi]_{-1}=-iY^3[(\dot\rho^a)^*\bar\psi_a]_{-1}=-iY^3(\dot\rho^a)^*\bar\sigma_a\ .
\eea
On the other hand, we have that
\bea
(F^{(0)}+i\d\calt)_{-1}=\frac{i}{2}\calg^{ab}\frac{\langle\sigma_b,F^{(0)}+i\d\calt\rangle}{\d\text{vol}_0}\,\bar\sigma_a\ ,
\eea
where $\calg^{ab}$ is the inverse of $\calg_{ab}$. We can now use the following identities
\begin{equation}
	\begin{split}
		\langle \chi,1\rangle&= -\frac{i}{3!}\cali_{abc}\rho^a\rho^b\rho^c\,\d\text{vol}_0\ ,\\
		\langle \chi,\omega_a\rangle&= \frac{1}{2}\cali_{abc}\rho^b\rho^c\,\d\text{vol}_0\ ,\\
		\langle \chi,\tilde\omega^a\rangle&= 
		i\rho^a\,\d\text{vol}_0\ ,\\
		\langle \chi,\d\text{Vol}_0\rangle&= -\,\d\text{vol}_0\ ,
	\end{split}
\end{equation}
and
\begin{equation}
	\begin{split}
		\langle \psi_a,1\rangle&=-\frac12\cali_{abc}\rho^b\rho^c\,\d\text{vol}_0=-i\frac{\del}{\del \rho^a}\langle \chi,1\rangle\ ,\\
		\langle \psi_a,\omega_b\rangle&=-\frac{i}{2}\cali_{abc}\rho^c\,\d\text{vol}_0=-i\frac{\del}{\del \rho^a}\langle \chi,\omega_b\rangle\ ,\\
		\langle \psi_a,\tilde\omega^b\rangle&=\delta_a^b\,\d\text{vol}_0=-i\frac{\del}{\del \rho^a}\langle \chi,\tilde\omega^b\rangle\ ,\\
		\langle \psi_a,\d\text{vol}_0\rangle&=0\ ,
	\end{split}
\end{equation}
to compute 
\begin{equation}
	\begin{split}
	\langle\sigma_a,F^{(0)}+i\d\calt\rangle&=\langle\psi_a,F^{(0)}+i\d\calt\rangle+\frac{i}{2}\,\frac{\cali_a}{\cali}\langle\chi,F^{(0)}+i\d\calt\rangle\\
		&=-i\frac{\del}{\del \rho^a}\langle\chi,F^{(0)}+i\d\calt\rangle+\frac{i}{2}\,\frac{\cali_a}{\cali}\langle\chi,F^{(0)}+i\d\calt\rangle \\
		&= \frac{4i}{M^3_{\rm P}V_0}\Big[\frac{\del W}{\del \rho^a}-\frac12\,\frac{\cali_a}{\cali} W\Big]\,\d\text{vol}_0\\
		&=  \frac{4i}{M^3_{\rm P}V_0}D_a W\,\d\text{vol}_0\ ,
	\end{split}
\end{equation}
where 
\bea
D_a W\equiv \frac{\del W}{\del \rho^a}+\frac{\del \calk}{\del \rho^a}\,W\ ,
\eea
and $W$ and $\calk$ are given in (\ref{cosetsup}) and (\ref{cosetkaehler}). Then,  we can write
\bea
(F^{(0)}+i\d\calt)_{-1}=-\frac{2}{V_0 M^3_{\rm P}}\,\calg^{ab}D_a W\,\bar\sigma_b \ .
\eea
Taking into account that $*_\bindex\sigma_a=-i\sigma_a$ and
\bea
G_{a\bar b}\equiv \frac{\del^2 \calk}{\del \rho^a\del\bar\rho^b}=-\frac{1}{4\cali}\calg_{ab}\ ,
\eea
we obtain
\bea
(\dot\rho^a)^*=\frac{Y^{-3}e^{4\warp}}{M^3_{\rm P}V_0\cali}\,e^{i\vartheta}G^{\bar{a}b}D_b W\ .
\eea
Noticing that 
\bea\label{warpid}
e^{4\warp}=|Y|^4\frac{(i\langle\calz^0,\bar\calz^0\rangle)^{2/3}}{(i\langle T,\bar T\rangle)^{2/3}}=|Y|^4V_0\cali e^{\calk/3}\ ,
\eea
and using (\ref{eingauge}), we obtain \eqref{dotrho}.

Let us now turn to (\ref{holflow1}), again considering its complex conjugate equation for convenience. In order to proceed, we need some preliminary results. First observe that 
\bea\label{identity}
\calk_{IJ}=\frac{1}{\text{Vol}_0\,\calh}\int_{\calm_6}\langle *_\bindex\alpha_I,\alpha_J\rangle\ ,
\eea 
where, $\calk_{IJ}=\del^2\calk/\del t^I\del t^J$. In this section, we generically use this convention to write derivatives with respect to $t^I$. In order to prove (\ref{identity}), let us first observe that
\begin{equation}
	\begin{split}
		\calk_I&=-\frac{2}{\calh}\,\calh_I= -\frac{1}{{\rm Vol}_0\calh}\,\int_{\calm_6}\langle \alpha_I,\Im T\rangle,\\
		\calk_{IJ}&= -\frac{2}{\calh}\Big(\calh_{IJ}-\frac{1}{\calh}\,\calh_I\calh_J\Big)\ .
	\end{split}
\end{equation}
Then, in order to compute $\calh_{IJ}$, one can use the decomposition of a three--form in its components in $V_{3}\oplus V_1\oplus V_{-1}\oplus V_{-3}$ (where the $V_k$ were defined in (\ref{eq:vk})). One can prove that \cite{hitchin}
\bea
\calh_{IJ}=\frac{1}{2\text{Vol}_0}\int_{\calm_6}\langle \alpha_I,J^{\rm Hit}\cdot\alpha_J\rangle\ ,
\eea
where $J^{\rm Hit}$ is an almost complex structure defined as follows: it takes value $-i$ when it acts on $V_3\oplus V_1$ and $i$ when it acts on $V_{-1}\oplus V_{-3}$. Then, by using the fact that  $*_\bindex=i$ when it acts on $V_3\oplus V_{-1}$ and $*_\bindex=-i$ when it acts on $V_{1}\oplus V_{-3}$ and taking into account that $\langle\alpha_I,\Re T\rangle=0$, it is easy to see that (\ref{identity}) is indeed valid.
This implies that
\bea \label{starbeta}
*_\bindex\beta^I=\frac{1}{\calh}\,\calk^{IJ}\alpha_J\ .
\eea
Furthermore, observe that from the homogeneity of $\calh$ it follows that
\bea\label{tk}
t^I=-\calk^{IJ}\calk_J\ .
\eea

Coming back to the complex conjugate of (\ref{holflow1}), on the l.h.s.\ we have
\bea\label{dertau}
\partial_r\bar\calt=(\dot \tau^I)^*\alpha_I\ ,
\eea
while on the r.h.s.\ the following quantities appear:
\bea\label{derw}
e^{i\vartheta}(\d\calz)_0=iY^3e^{i\vartheta}\rho^a q_{a,I}\beta^I\ , \qquad \Re T\,\frac{W_{\rm sc}}{N}=\frac{1}{M^2_{\rm P}}\,e^{\calk/2+i\vartheta}W t^I\alpha_I\ .
\eea
Then, taking into account that in the conventions of this section the K\"ahler metric is given by $G_{I\bar J}=\calk_{IJ}/4$, and that $Y$ is fixed to be (\ref{eingauge}), from (\ref{derw}), (\ref{warpid}) and \eqref{starbeta} we find that  the r.h.s.\  of the complex conjugate of (\ref{holflow1}) is given by
\begin{equation}
\begin{split}
	&\Big(-\frac{V_0 M_{\rm P}}{4} \rho^a q_{a,I}+\frac{W}{M^2_{\rm P}}\frac{\partial\calk}{\del \tau^I}\Big)e^{\calk/2+i\vartheta}G^{I\bar J}\alpha_J\\
	&= \frac{1}{M^2_{\rm P}}e^{\calk/2+i\vartheta}\Big(\frac{\partial W}{\del \tau^I}   +W\frac{\partial\calk}{\del \tau^I}\Big)G^{I\bar J}\alpha_J\ , \label{lasteq}
\end{split}	
\end{equation}
where we have used the explicit expression of $W$ given in (\ref{cosetsup}). By plugging (\ref{lasteq}) and (\ref{dertau}) into the complex conjugate of (\ref{holflow1}) one finally gets the complex conjugate of \eqref{dottau}.

\end{appendix}


\end{document}